\documentclass[12pt,letterpaper]{article}
\pdfoutput=1
\usepackage{graphicx,array}
\usepackage{color}
\usepackage{latexsym}
\usepackage{amsthm}
\usepackage{amsmath}
\usepackage{amssymb}
\usepackage{slashed}

\usepackage{float}
\usepackage{url}
\usepackage{mathrsfs}
\usepackage{caption}

\usepackage{dsfont}
\usepackage{verbatim}
\usepackage{epsfig}
\usepackage{slashed}
\usepackage{bbold}
\usepackage{psfrag}
\usepackage[svgnames]{xcolor}
\PassOptionsToPackage{caption=false}{subfig}
\usepackage{subcaption}
\usepackage{xfrac}
\usepackage{multirow}
\usepackage{booktabs}
\usepackage[colorlinks=true,linkcolor=MediumBlue,citecolor=Green,urlcolor=violet]{hyperref}
\usepackage{cite}
\usepackage[normalem]{ulem}
\usepackage{makecell}

\newcommand{\be}{\begin{equation}}
\newcommand{\ee}{\end{equation}}
\newcommand{\bea}{\begin{eqnarray}}
\newcommand{\eea}{\end{eqnarray}}

\newcommand{\mttcut}{m_{t\bar t}^{\textrm{max}}}

\newcommand{\rmZ}{{\textrm{Z}}}

\def\({\left(}
\def\){\right)}

\usepackage{soul}

\begin{document}
\vspace{.6cm}
\begin{center}
{\LARGE \bf  New Physics from High Energy Tops}
\bigskip\vspace{1cm}{

Marco Farina$^1$, Cristina Mondino$^2$, Duccio Pappadopulo$^{2,3}$, and Joshua T. Ruderman$^2$ }
\\[7mm]
 {\it \small
$^1$C.N.Yang Institute for Theoretical Physics, Stony Brook, NY 11794, USA \\
$^{2}$Center for Cosmology and Particle Physics, Department of Physics, \\ 
New York University, New York, NY 10003, USA\\
$^{3}$Bloomberg LP, New York, NY 10022, USA\\
 }

\end{center}

\bigskip \bigskip \bigskip \bigskip

\vspace{-1.0cm}\centerline{\bf Abstract} 

\begin{quote}
Precision measurements of high energy top quarks at the LHC constitute a powerful probe of new physics. 
We study the effect of four fermion operators involving two tops and two light quarks on the high energy tail of the $t\bar t$ invariant mass distribution. 
We use existing measurements at a center of mass energy of 13\,TeV, and state of the art calculations of the Standard Model contribution, to derive bounds on the coefficients of these operators. We estimate the projected reach of the LHC at higher luminosities and discuss the validity of these limits within the Effective Field Theory description.  We find that current measurements constrain the mass scale of these operators to be larger than about 1-2\,TeV, while we project that future LHC data will be sensitive to mass scales of about 3-4\,TeV\@.  We apply our bounds to constrain composite Higgs models with partial compositeness and models with approximate flavor symmetries. We find our limits to be most relevant to flavor non-universal models with a moderately large coupling of the heavy new physics states to third generation quarks. 

\end{quote}

\vfill
\noindent\line(1,0){188}
{{\scriptsize{ \\  \texttt{$^1$ \href{mailto:farina.phys@gmail.com}{farina.phys@gmail.com}\\ \texttt{$^2$ \href{mailto:cri.mondino@gmail.com}{cri.mondino@gmail.com}\\ $^3$ \href{mailto:duccio.pappadopulo@gmail.com}{duccio.pappadopulo@gmail.com}\\ $^4$ \href{mailto:ruderman@nyu.edu}{ruderman@nyu.edu}}}}}
\newpage

\section{Introduction}
\label{sec:intro}

Of all particles in the Standard Model (SM), the top quark is the one that couples most strongly to the Higgs boson. As such it is the particle that most severely contributes to the Higgs hierarchy problem. For this reason, natural extensions of the SM generically predict modifications of the Higgs and top sectors of the theory, either in the form of new weakly coupled states or new strong dynamics. 

On the other hand, all measurements performed at the Large Hadron Collider (LHC) seem to show agreement with the predictions of the SM\@. 

With no decisive indication of New Physics (NP) emerging from the data, a promising way to organize the available results is provided by the SM Effective Field Theory~\cite{Buchmuller:1985jz,Feruglio:1992wf,Grzadkowski:2010es,Alonso:2012px,Contino:2013kra,Davidson:2013fxa,Biekoetter:2014jwa,deVries:2014apa,Falkowski:2015wza}. The effects of new particles and phenomena that are too heavy to be directly accessed at the LHC can be described in full generality by adding operators of dimension larger than 4 to the SM Lagrangian,
\be
\mathscr L = \mathscr L_{SM}+\sum_i C^{(5)}_i O^{(5)}_i+\sum_i C^{(6)}_i  O^{(6)}_i+\ldots \, \, .
\ee

Partly motivated by the special role the top quark has in relation to the hierarchy problem, we consider operators modifying the high energy tails of kinematic distributions arising from $t\bar t$ pair production at the LHC\@. Similar shape analyses for tails of other differential distributions have been performed by Refs.~\cite{Cirigliano:2012ab, deBlas:2013qqa,Falkowski:2016cxu,Farina:2016rws,Raj:2016aky,Alioli:2017jdo,Greljo:2017vvb,Bellazzini:2017bkb,Azatov:2017kzw,Panico:2017frx,Franceschini:2017xkh,Alioli:2017nzr}. We focus on the $t\bar t$ invariant mass distribution and we consider those operators for which the leading order effect at high energies comes from the interference between the QCD SM amplitude and the amplitude generated by a single insertion of a dimension 6 operator.  We also require such corrections to be nonvanishing in the limit in which all SM masses are much smaller than the typical energy scale of the process that is considered.

These requirements single out the set of gauge invariant dimension six operators shown in Table~(\ref{tab4f}). Using Lorentz, $SU(3)_c$, and $SU(2)_{EW}$ Fierz identities all the operators can be written as four fermion operators involving the product of two color octet currents: a $t\bar t$ one and a light quark one. This is indeed the same structure of the $q\bar q\to t\bar t$ amplitude in the SM\@. The full set of four fermion operators contributing to the $pp\to t\bar t$ cross section is shown in Appendix~\ref{operators}.

Other groups have studied the impact of precise measurements of top quark observables on the SM EFT~\cite{AguilarSaavedra:2010zi,AguilarSaavedra:2011vw,Kamenik:2011dk,Zhang:2012cd,Rontsch:2014cca,Degrande:2014tta,Durieux:2014xla,Rontsch:2015una,Franzosi:2015osa,Buckley:2015nca,Dror:2015nkp,Buckley:2015lku,Zhang:2016omx,Bylund:2016phk,Schulze:2016qas,Cirigliano:2016nyn,Maltoni:2016yxb,Englert:2017dev,Zhang:2017mls,AguilarSaavedra:2018nen,Chala:2018agk}.  Here we focus on the most recent data measuring the $t\bar t$ differential production cross section, and state of the art theoretical calculations, to extract reliable bounds on the dimension 6 operators appearing in Table~(\ref{tab4f}).

\begin{table}[t]
\caption*{95\%~CL bounds on operator coefficients ($\times10^{3}$) at $\sqrt s=13$\,TeV\vspace{-0.3cm}}
\begin{center}
{\small
\begin{tabular}{c|c|c|c|c} 
  & Operator & \makecell{35.8\,fb$^{-1}$\,(CMS) \\  {\footnotesize Observed $\ \ $ Expected}} & 300\,fb$^{-1}$&3\,ab$^{-1}$ \\ \hline \hline 
$ O_{Qq}^{(3)}$ & $\bar Q\gamma^\mu T^A\tau^aQ\, \bar q\gamma_\mu T^A\tau^aq$& $[-25,19]\ \ \ \ [-27,23]$ &$[-6.9,5.5]$ &$[-5.4,4.1]$\\
$O_{Qq}$ & $\bar Q\gamma^\mu T^AQ\, \bar q\gamma_\mu T^Aq$&$[-32,12]\ \ \ \ [-32,19]$&$[-7.1,5.2]$&$[-5.3,4.1]$\\
$O_{Qu}$ & $\bar Q\gamma^\mu T^AQ\, \bar u\gamma_\mu T^Au$&$[-37,17]\ \ \ \ [-39,24]$&$[-7.6,5.8]$&$[-5.5,4.2]$\\ 
$O_{Qd}$ & $\bar Q\gamma^\mu T^AQ\, \bar d\gamma_\mu T^Ad$&$[-46,26]\ \ \ \ [-49,36]$&$[-15.,13.]$&$[-13.,11.]$ \\ 
$O_{Uq}$ & $\bar U\gamma^\mu T^AU\, \bar q\gamma_\mu T^Aq$&$[-32,12]\ \ \ \ [-32,19]$&$[-7.1,5.2]$& $[-5.3,4.1]$\\ 
$O_{Uu}$ & $\bar U\gamma^\mu T^AU\, \bar u\gamma_\mu T^Au$&$[-37,17]\ \ \ \ [-39, 25]$&$[-7.6,5.8]$&$[-5.5,4.2]$ \\
$O_{Ud}$ & $\bar U\gamma^\mu T^AU\, \bar d\gamma_\mu T^Ad$&$[-46,26]\ \ \ \ [-49, 36]$&$[-15.,13.]$&$[-13.,11.] $\\
\end{tabular}
} 
\vspace{0.3cm}
\caption{\label{tab4f}\footnotesize Gauge and Lorentz structure of dimension 6 four fermion operators leading to nonvanishing interference with the SM QCD $q\bar q\to t\bar t$ amplitude at leading order and neglecting quark masses. We use capital letters $Q$ and $U$ to denote the third generation quark doublet and up-type singlet, while lowercase $q, u$, and $d$ denote quarks from the first two generations. $SU(3)_c$ and $SU(2)_{EW}$ generators are denoted by $T^A$ and $\tau^a$. In all the operators we sum over light quark flavors.  We report 95\%\,CL bounds on $c_i$, where the operators,  $\mathcal{O}_i$, are normalized to have coefficients $c_i g_s^2 / m_t^2 \, \mathcal{O}_i$.  Current bounds are extracted from CMS data~\cite{Sirunyan:2018wem} (both observed and expected), while projections correspond to the higher luminosities of 300 and 3000~fb$^{-1}$ at $\sqrt s = 13$\,TeV.  
}
\end{center}
\end{table}


This paper is organized as follows. In Sec.~\ref{sec:precision} we describe the theory calculations that are available for the $t\bar t$ invariant mass distribution and their uncertainties. We describe the experimental measurements and the statistical methods that we use to extract bounds and future projections. In Sec.~\ref{sec:bounds} bounds on a set of dimension 6 four fermion operators are presented and their validity in the framework of the Effective Field Theory is discussed. In Sec.~\ref{sec:implications} we discuss the implications of such bounds on relevant NP models such as composite Higgs models and flavor models with $U(3)$ or $U(2)$ flavor symmetries. In Sec.~\ref{sec:conclusions} we present our conclusions.


\section{Precision measurements in high energy $t\bar t$ observables}
\label{sec:precision}

The differential cross section of top quark pair production at the LHC is one of the most accurately known hadronic observables. This is due to the groundbreaking work of~Refs.~\cite{Czakon:2013goa, Czakon1,Czakon2,Czakon3,Czakon4},  which achieve full NNLO QCD and  NLO EW accuracy for (undecayed) final state  top quarks.

From the experimental side both CMS~\cite{Sirunyan:2018wem} and ATLAS~\cite{Aaboud:2017fha} provide measurements of the differential $t\bar t$ cross section in the lepton plus jets final state and ATLAS in the fully hadronic final state~\cite{Aaboud:2018eqg} at 13\,TeV center of mass energy. In this paper we use the CMS result, which uses a luminosity of 35.8\,fb$^{-1}$. The differential NNLO predictions of Ref.~\cite{Czakon4} are not available for the kinematic cuts of ATLAS~\cite{Aaboud:2018eqg}, while Ref.~\cite{Aaboud:2017fha} does not provide unfolded values for the parton level cross section. Previous measurements of the differential $t\bar t$ cross section have been performed by ATLAS~\cite{Aad:2012hg,Aad:2014zka,Aad:2015eia,Aad:2015hna,Aad:2015mbv,Aaboud:2016iot,Aaboud:2016syx} and CMS~\cite{Chatrchyan:2012saa,Khachatryan:2015oqa,Khachatryan:2016gxp,Khachatryan:2016mnb,Sirunyan:2017azo}.  While we use differential cross section measurements to look for smooth effects parameterized by the Effective Field Theory, we note that top pair production measurements have also been used to search for sharp resonances by ATLAS~\cite{Aad:2015fna,Aaboud:2017hnm,Aaboud:2018mjh} and CMS~\cite{Khachatryan:2015sma,Sirunyan:2017uhk,Sirunyan:2018ryr}. The implications of using NNLO QCD theoretical predictions for such bump hunts in the $t\bar t$ invariant mass spectrum was studied by Ref.~\cite{Czakon:2016vfr}.

In the left panel of Fig.~(\ref{CMSvstheory}) we compare the measurement of Ref.~\cite{Sirunyan:2018wem}, of the unfolded parton level $t\bar t$ invariant mass distribution, with the theory calculation from Ref.~\cite{Czakon4}. We include experimental uncertainties and their correlations from Refs.~\cite{Sirunyan:2018wem}. Theory uncertainties, including QCD scale variation and PDF uncertainties, are taken from Ref.~\cite{Czakon4} in which  PDF uncertainties are calculated using the {\texttt{PDF4LHC15}}~\cite{Butterworth:2015oua} set extended with {\texttt{LUXqed}}~\cite{Manohar:2016nzj}. This PDF set includes a combination of the results from Refs.~\cite{Ball:2014uwa,Harland-Lang:2014zoa,Dulat:2015mca}, where the only top observable included in the fits is the total $t\bar t$ production cross sections at 7 and 8\,TeV, which we do not expect to be significantly contaminated by the EFT operators that we consider below, which produce effects that grow with energy.  We take scale uncertainties and PDF uncertainties to be uncorrelated from each other.  On the right panel of Fig.~(\ref{CMSvstheory}), we show the relative size of experimental and theory uncertainties. The largest source of uncertainty is experimental systematics, which is as large as 20\% in the last invariant mass bin.  Note that CMS measures the cross section times branching fraction of semileptonic $t\bar t$ events, $\sigma_{t \bar t} \times \textrm{BR}_l$, where at parton level $\textrm{BR}_l \approx 0.29$.\footnote{The partonic semileptonic branching fraction includes decays where one top decays to an electron or muon while the other top decays to hadrons (and neither top decays to tau leptons).  Therefore, $\textrm{BR}_l \approx 4 \, BR_{W \rightarrow l \nu} (1-3 \, \textrm{BR}_{W \rightarrow l \nu}) \approx 0.29$ where $\textrm{BR}_{W \rightarrow l \nu} \approx 0.109$~\cite{Tanabashi:2018oca}.}

Goodness of fit is evaluated by constructing a $\chi^2$ statistic,
\be\label{chi2fit}
\chi^2 = \sum_{I,J}({\textrm{th}}^{(I)} - {\textrm{exp}}^{(I)})\left(\Sigma^{-1}\right)_{I,J}({\textrm{th}}^{(J)} - {\textrm{exp}}^{(J)}) \, \, ,
\ee
where ${\textrm{th}}^{(I)}$ and ${\textrm{exp}}^{(I)}$ are the experimental and theory prediction in the $I$-th $t\bar t$ invariant mass bin, and $\Sigma$ is the total covariance matrix including all uncertainties described above. Assuming the usual asymptotic behavior of $\chi^2$ we can associate a p-value to the SM fit,
\be
{\textrm{p}} = 1- {\textrm{cdf}}_{\chi^2_{n}}(\chi^2) \, \, ,
\ee
where ${\textrm{cdf}}_{\chi^2_{n}}$ is the cumulative chi-squared distribution with $n=10$ degrees of freedom. The p-value we obtain for the fit is decent, ${\textrm{p}}=0.10$, which we take as an indication that both the uncertainties and the theory prediction are under control.

\begin{figure}[t]
\begin{center}
~~\includegraphics[width=0.475\textwidth]{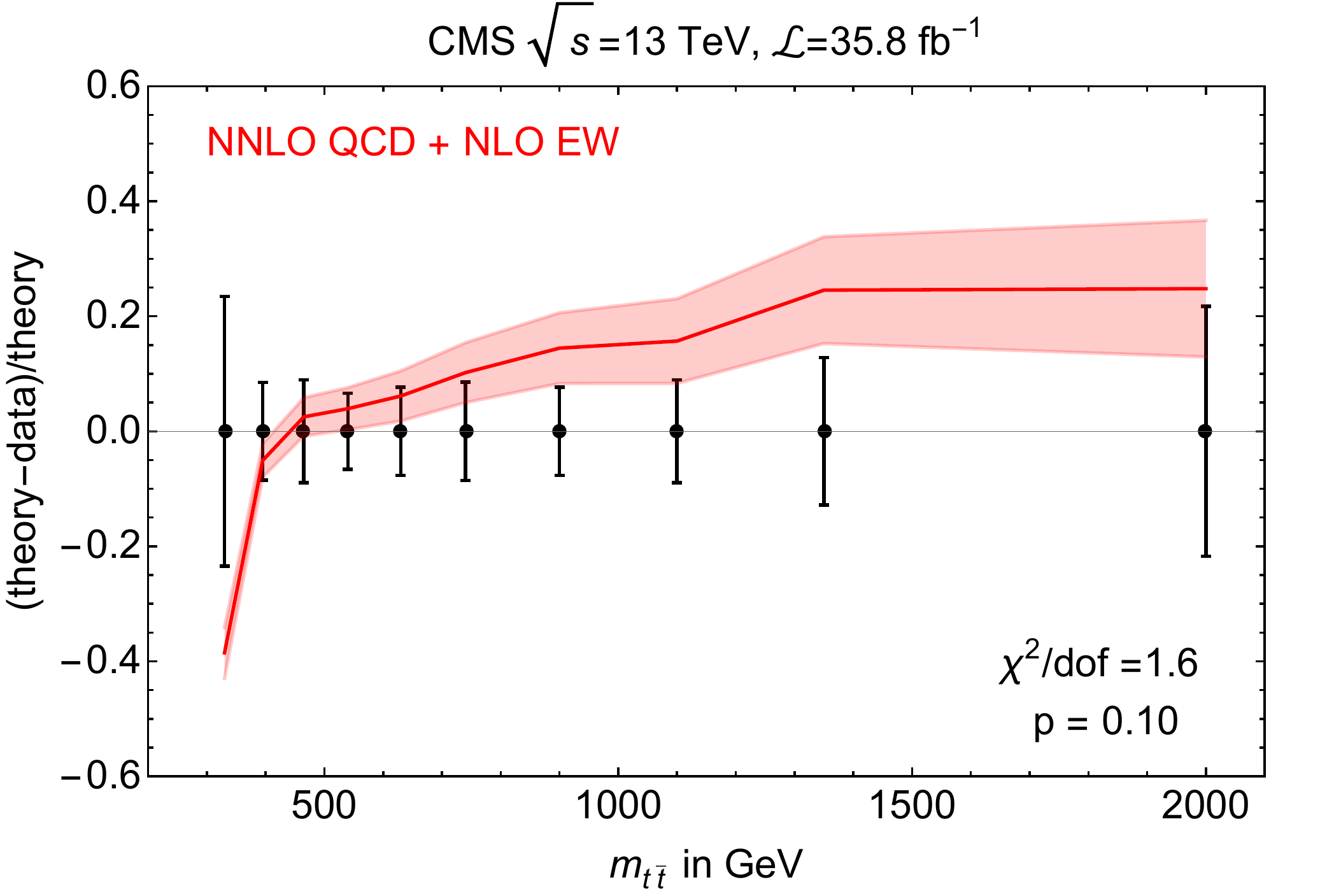}~~~~~\includegraphics[width=0.475\textwidth]{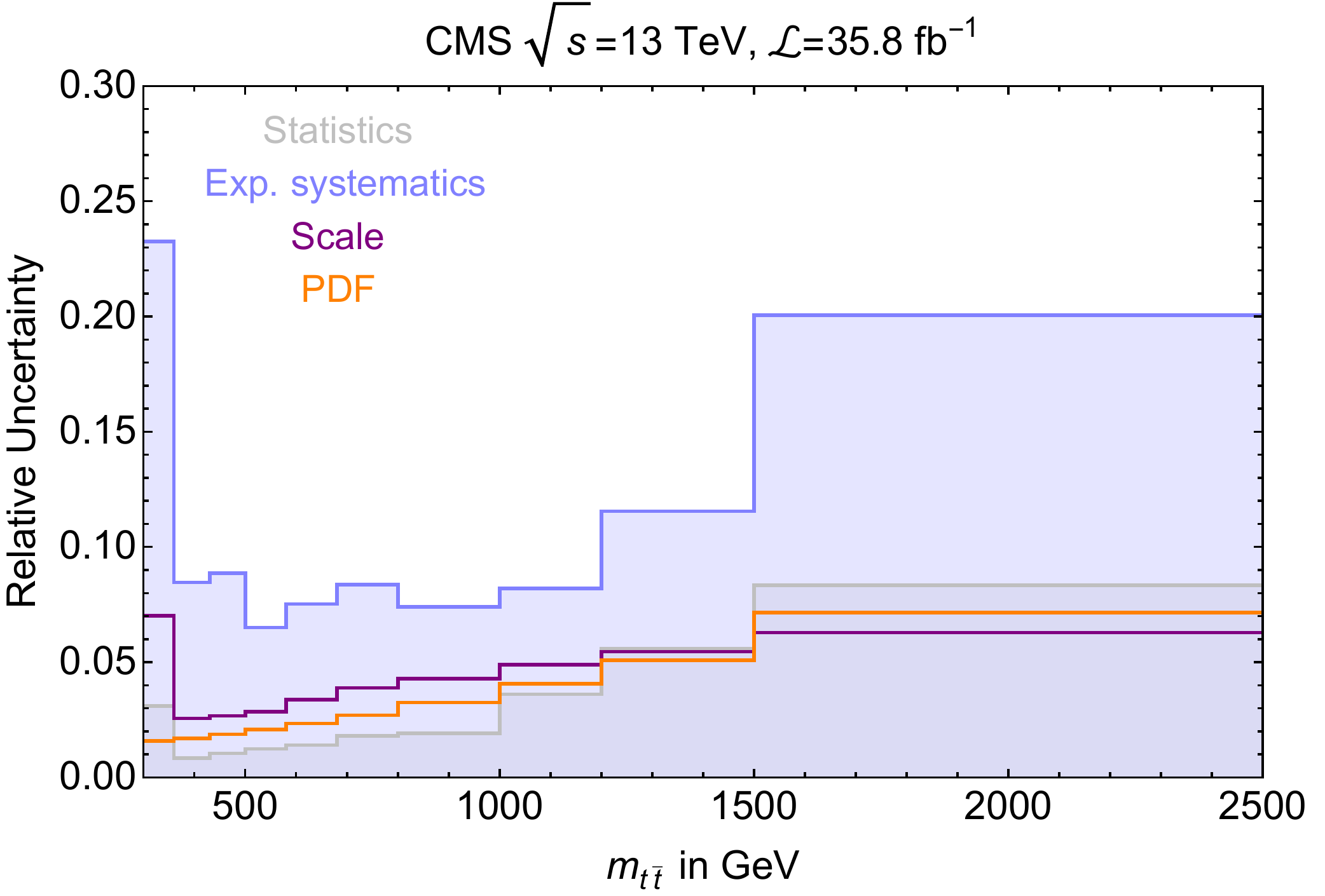}
\end{center}
\vspace{-.3cm}
\caption{ \footnotesize
\emph{Left}: comparison of theory prediction to experimental data from Ref.~\cite{Sirunyan:2018wem}. \emph{Right}: summary of theory uncertainties from Ref.~\cite{Czakon4} and experimental uncertainties from Ref.~\cite{Sirunyan:2018wem}. The uncertainties on both plots are $1 \sigma$.
}
\label{CMSvstheory}
\end{figure}

In order to make projections for future measurements of the $t\bar t$ invariant mass distribution, that will benefit from more luminosity and therefore higher statistics at higher energies, we extend the invariant mass range until $m_{t \bar t} = 6$\,TeV\@.
We write the full covariance matrix for the uncertainties as
\be\label{covproj}
\Sigma = \Sigma_{\textrm{theory}}+\Sigma_{\textrm{stat}}+\Sigma_{\textrm{syst}} \, \, .
\ee
Theory uncertainties and correlations, $\Sigma_{\textrm{theory}}$, including scale variation and PDF uncertainties, are evaluated in the new mass range using Ref.~\cite{Czakon4}, as shown in Fig.~(\ref{errorproj}). 
For the statistical uncertainty contribution to the full covariance, $\Sigma_{\textrm{stat}}$, we use the Gaussian limit,
\be
\left(\Sigma_{\textrm{stat}}\right)_{I,J} = \frac{\sigma^{(I)}}{{\textrm{BR}_l}\times\epsilon\times\mathcal{L}}\,\delta_{IJ} \, \, ,
\ee
where as above $\textrm{BR}_l = 0.29$.  The current measurement of Ref.~\cite{Sirunyan:2018wem} has an overall selection efficiency of about 4 and 5\% at the parton and particle levels, respectively. For our future projections we take $\epsilon=0.05$, independent of the invariant mass.

Experimental systematic uncertainties are modeled by including two fractional sources of uncertainty,
\be
\left(\Sigma_{\textrm{syst}}\right)_{I,J} =(\delta_C^2+\delta_U^2\delta_{IJ})\sigma^{(I)}\sigma^{(J)} \, \, ,
\ee
with $\delta_C$ being completely correlated and $\delta_U$ fully uncorrelated. We choose $\delta_C=\delta_{U}=7\%$ to roughy match current experimental uncertainties~\cite{Sirunyan:2018wem}.\footnote{We validated this simplified treatment of experimental uncertainties, for future projections, by verifying that we produce similar bounds on operators from the CMS measurement~\cite{Sirunyan:2018wem} when using $\delta_C=\delta_{U}=7\%$ and when using the full experimental covariance matrix.}

\begin{figure}[t]
\begin{center}
~~\includegraphics[width=0.475\textwidth]{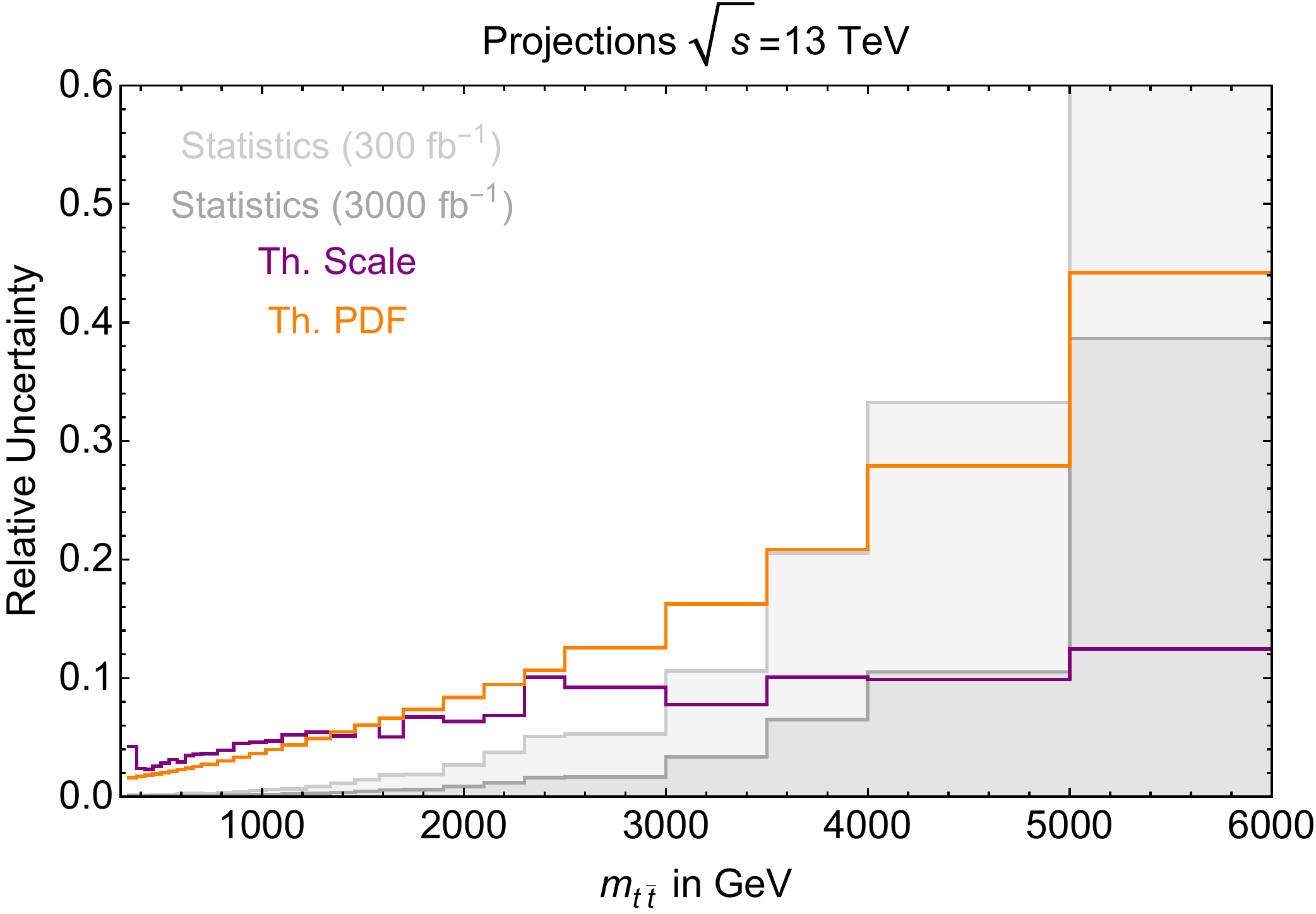}~~~~~\includegraphics[width=0.375\textwidth]{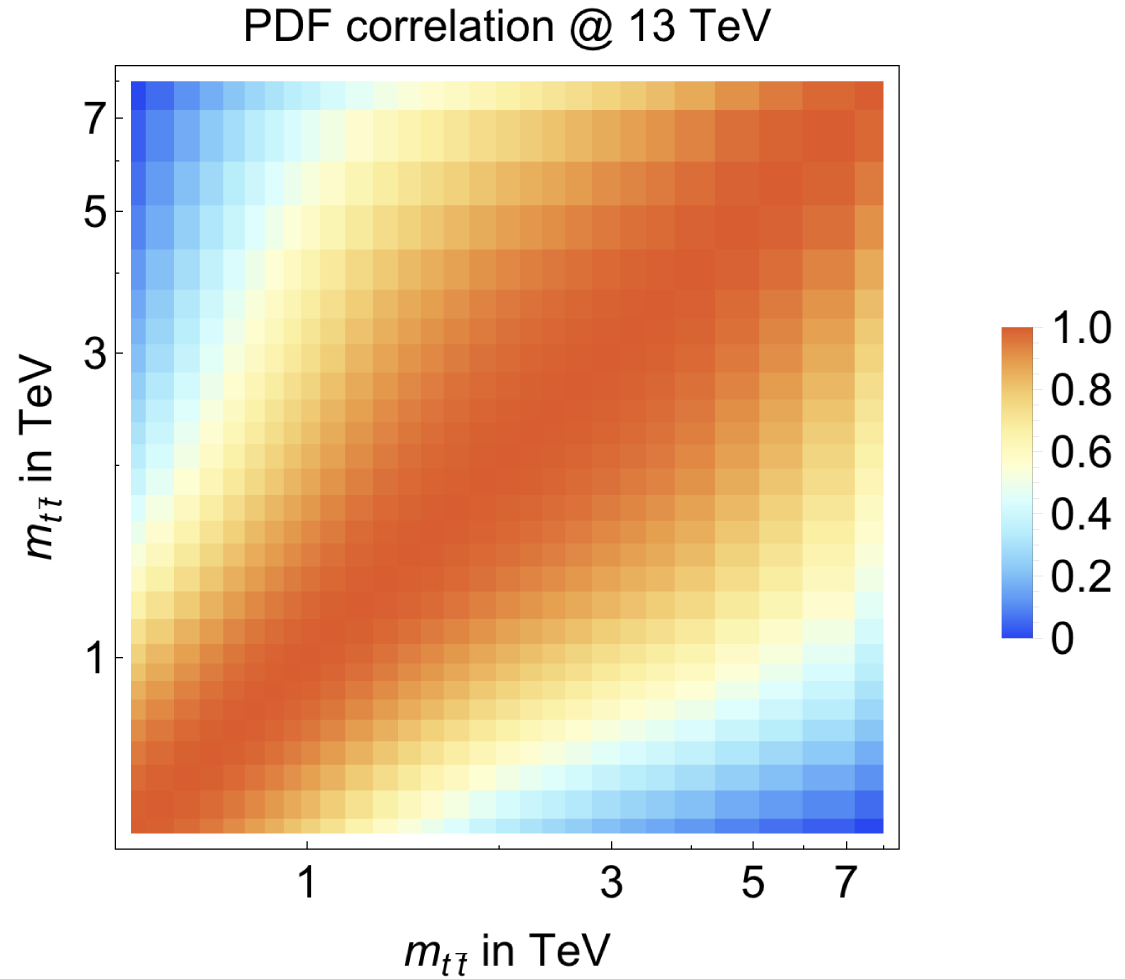}
\end{center}
\vspace{-.3cm}
\caption{ \footnotesize
Projected theory and statistical uncertainties ($1 \sigma$) for the $t\bar t$ invariant mass distribution at the 13\,TeV LHC\@. Statistical uncertainties are evaluated for the two different integrated luminosities of 0.3 and 3\,ab$^{-1}$.  The left panel shows the size of the uncertainty in each invariant mass bin, and the right panel shows the correlation of the PDF uncertainty across different invariant masses.
}
\label{errorproj}
\end{figure}

\section{Bounds}\label{sec:bounds}

We are now ready to derive the bounds shown in Table~(\ref{tab4f}). To do so we normalize the operators according to
\be\label{deltaL4f}
\mathscr L\supset \sum_i \frac{g_s^2 c_i}{m_t^2} O_i \, \, ,
\ee
where the sum is over the operators in Table~(\ref{tab4f}), $g_s=1.22$ is the strong coupling constant taken here as a fixed reference value, and $m_t=173.3$\,GeV is the top quark mass. 
The value of the $ t\bar t$ cross section, $\sigma^{(I)}$, integrated over a range of invariant masses, $I$, is a quadratic polynomial in the coefficients $c_i$,
\be\label{sigmaeft}
\sigma^{(I)} = \sigma^{(I)}_{SM}+ \sum_i c_i  \sigma^{(I)}_{i}+  \sum_{i,j} c_i c_j  \sigma^{(I)}_{i,j} \, \, .
\ee 
The linear term corresponds to interference between the SM amplitude and the NP one, and the quadratic terms are due to the square of the NP amplitude. The numerical values of these terms are obtained at leading order by integrating the squared amplitudes, shown in Appendix~(\ref{operators}), over the relevant phase space.

\begin{figure}[t]
\begin{center}
\includegraphics[width=0.475\textwidth]{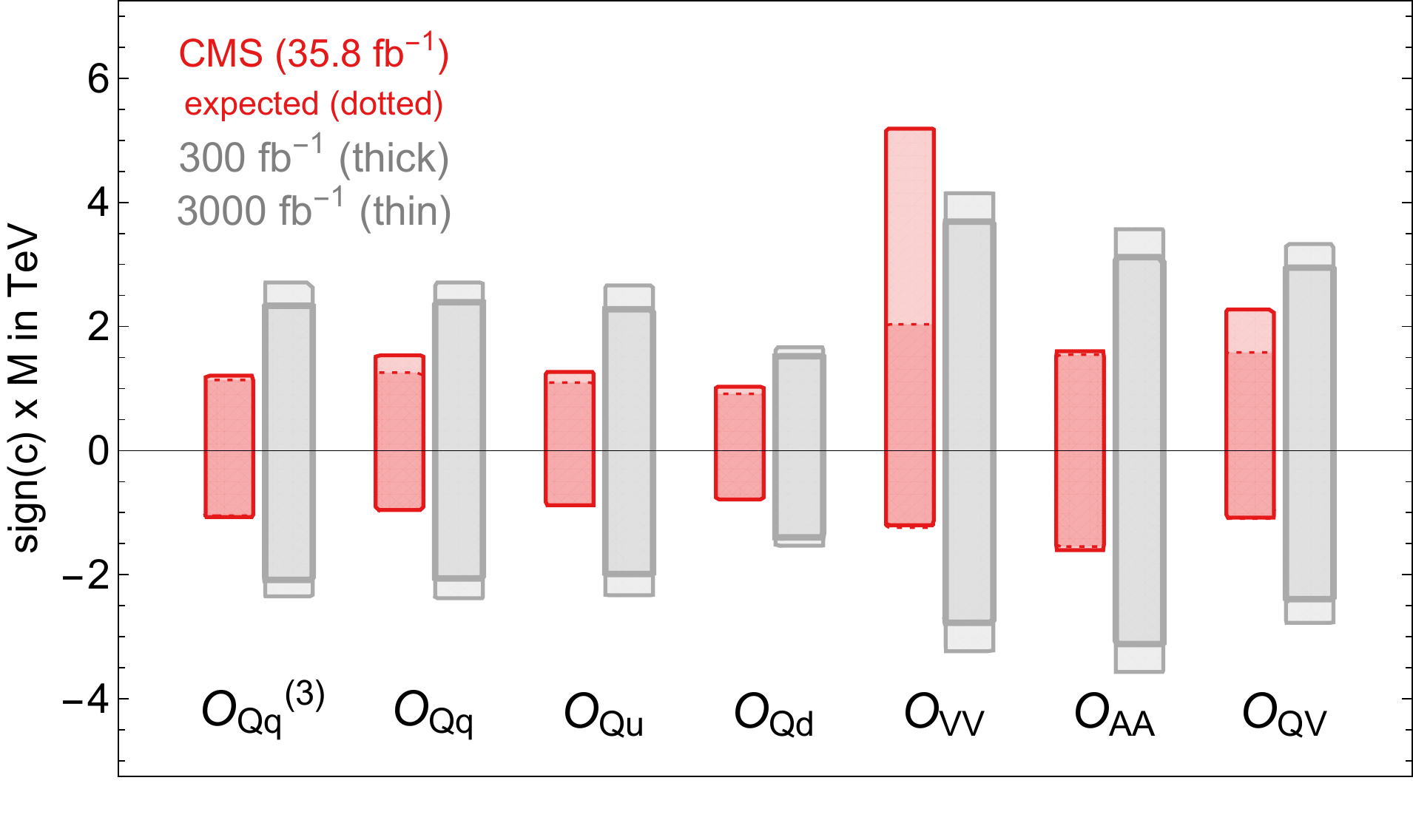}~~~~~~\includegraphics[width=0.475\textwidth]{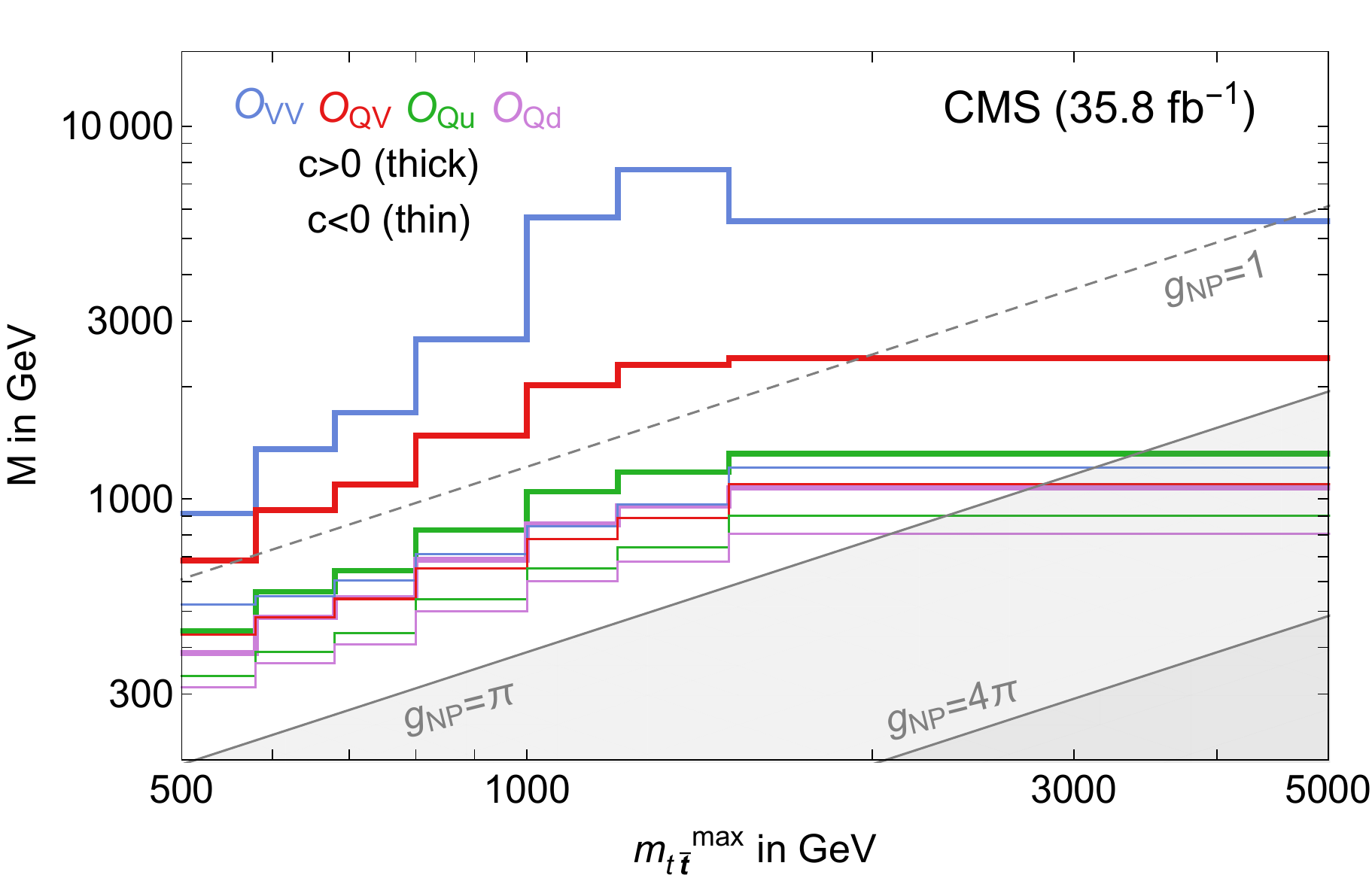}
\end{center}
\vspace{-.3cm}
\caption{ \footnotesize
\emph{Left}: 95\% CL limits on various dimension 6 operators, in terms of a reference mass scale $M\equiv m_t/\sqrt{|c|}$ (see Eq.~(\ref{deltaL4f})), from CMS data (observed and expected bounds) and our high luminosity projection study of the $t\bar t$ invariant mass distribution.  For each operator, the shaded region is excluded. \emph{Right}: We show how the bounds degrade by only including in the fit those events for which the reconstructed $t\bar t$ invariant mass is below a certain value $\mttcut$. Thick (thin) lines correspond to positive (negative) operator coefficient which in turn corresponds to constructive (destructive) interference with the SM\@.
}
\label{barplot}
\end{figure}

In order to define confidence intervals for the coefficients of the operators in Eq.~(\ref{deltaL4f}) using CMS data, we assume Gaussian uncertainties and construct the following statistic,
\be\label{chi2bounds}
\chi^2({\bf c}) = \sum_{I,J}({\sigma}^{(I)}({\bf c}) - {\sigma}^{(I)}_{\textrm{exp}})\left(\Sigma^{-1}\right)_{I,J}({\sigma}^{(J)}({\bf c}) - {\sigma}^{(J)}_{\textrm{exp}}) \, \, ,
\ee
where ${\bf c}$ is a subset of the coefficients $c_i$, ${\sigma}^{(I)}_{\textrm{exp}}$ are the cross section measurements, ${\sigma}^{(I)}({\bf c})$ their theory prediction, and $\Sigma$ is  as in Eq.~(\ref{chi2fit}).  
Defining ${\bf c}^* ={\textrm{argmin}}_{\bf c}~ \chi^2({\bf c})$, Wilks' theorem guarantees that the quantity $\Delta \chi^2({\bf c})\equiv \chi^2({\bf c}) -\chi^2({\bf c}^*)$ has a chi-squared distribution with number of degrees of freedom equal to the number of components of ${\bf c}$.

The 95\%\,CL intervals from the CMS measurement~\cite{Sirunyan:2018wem} are shown in Table~(\ref{tab4f}). We compare our results with other bounds present in the literature. Our limits are stronger than those derived by Ref.~\cite{Zhang:2017mls} using the 8\,TeV differential $m_{t\bar t}$ distribution measured by ATLAS~\cite{Aaboud:2016iot}. Notice however that Ref.~\cite{Zhang:2017mls} does not include experimental correlations which were not available. Another set of limits was obtained in the global fit of Refs.~\cite{Buckley:2015nca,Buckley:2015lku}, which include 8\,TeV differential $m_{t\bar t}$ distributions as an ingredient.  We note that Refs.~\cite{Buckley:2015nca,Buckley:2015lku} do not always include experimental covariances, and only include interference terms between the SM and NP, neglecting contributions that go as NP squared which, as we discuss below, can impact the bounds within the regime of validity of the EFT\@.  Four fermion operators have also been constrained using measurements of the charge asymmetry in top pair production, see for example Refs.~\cite{AguilarSaavedra:2011vw,Aguilar-Saavedra:2014kpa}. Ref.~\cite{Rosello:2015sck} uses the forward-backward asymmetry measured at Tevatron~\cite{Aaltonen:2012it,Abazov:2014cca}, and the charge asymmetry measured by CMS~\cite{Khachatryan:2015mna} and ATLAS~\cite{Aad:2015noh,Aad:2015lgx} at 8\,TeV to constrain four fermion operator coefficients.  When we consider the same linear combination of operators, our bounds are stronger.

 Projected bounds at higher luminosities are obtained by substituting ${\sigma}^{(I)}_{\textrm{exp}}$ with its expected SM value and the total covariance with the projected one, Eq.~(\ref{covproj}). The left panel of Fig.~(\ref{barplot}) displays these same bounds but in terms of an arbitrarily defined NP scale, 
\be\label{massscale}
M_i\equiv m_t/\sqrt{|c_i|}\, \, .
\ee 
Fig.~(\ref{barplot}) shows bounds on both individual operators and the following linear combinations of operators:
\begin{align}\nonumber
O_{VV}&\equiv (\bar Q\gamma^\mu T^AQ+\bar U\gamma^\mu T^AU)(\bar q\gamma_\mu T^Aq+\bar u\gamma_\mu T^Au+\bar d\gamma_\mu T^Ad)\,,\\\label{opscombo}
O_{AA}&\equiv (\bar Q\gamma^\mu T^AQ-\bar U\gamma^\mu T^AU)(\bar q\gamma_\mu T^Aq-\bar u\gamma_\mu T^Au-\bar d\gamma_\mu T^Ad)\,,\\\nonumber
O_{QV}&\equiv (\bar Q\gamma^\mu T^AQ)(\bar q\gamma_\mu T^Aq+\bar u\gamma_\mu T^Au+\bar d\gamma_\mu T^Ad)\, \, .
\end{align}

The effect of the smaller down quark PDF, versus the up quark, on the bounds can be readily observed by noticing that operators which only contribute to $d\bar d\to t\bar t$ display a weaker limit.  We note that the current CMS bound on the $O_{VV}$ operator shows the largest difference between the observed and expected limit.  This is because $O_{VV}$ has the largest interference with SM amplitude.  The CMS data are lower than expected, leading to a stronger than expected limit when the operator interferes constructively with the SM, which happens when $c_{VV}>0$.  When the operators interfere destructively, the current bounds are dominated by NP squared contribution, which is a steeper function of the NP scale and therefore less sensitive to fluctuations in the data.

In our study we only use information about the energy dependence of the $t\bar t$ cross section, but not its angular properties. This is because doubly-differential calculations of the $t\bar t$ cross sections are not yet available from Ref.~\cite{Czakon4}. This fact explains why the pairs of operators ($O_{Qq},O_{Uq}$), ($O_{Qu},O_{Uu}$), and ($O_{Qd},O_{Ud}$) have identical constraints.

In order to understand the validity of our bounds within the EFT framework we proceed as in Refs.~\cite{Farina:2016rws,Alioli:2017jdo}.  We introduce a variable $\mttcut$ and repeat the fit to the $t\bar t$ invariant mass distribution while conservatively only including bins characterized by a smaller invariant mass than this cutoff, $m_{t \bar t} < \mttcut$.
We show CMS bounds and projected ones as a function of $\mttcut$ in the right panel of Fig.~(\ref{barplot}) and in Fig.~(\ref{mttcut}) for a few representative combinations of the operators in Table~(\ref{tab4f}). The bounds are again expressed in term of the mass scale $M_i(\mttcut)\equiv m_t/\sqrt{|c_i|}$. We see that the strongest bound is for $O_{VV}$ and the weakest is for $O_{Qd}$. A negative sign of the operator coefficients implies destructive interference with the SM amplitude, leading to a weaker bound.

The form factor $\textrm{Z}$ studied in Ref.~\cite{Alioli:2017jdo} gives a contribution to $O_{VV}$
\be
-\frac{\textrm{Z}}{2 m_W^2}(D_\mu G^{\mu\nu A})^2\rightarrow -\frac{\rmZ g_s^2}{m_W^2} \,O_{VV} \, \, .
\ee
The projected bounds on $\textrm{Z}$ from dijet physics extracted by Ref.~\cite{Alioli:2017jdo} can then be directly compared to the bounds on the NP mass scale associated to $O_{VV}$. These bounds from dijets are also shown in Fig.~(\ref{mttcut}). We find that the $t\bar t$ invariant mass distribution is not competitive with dijet physics to constrain $\textrm{Z}$.

Among the operators in Table~(\ref{tab4f}), those involving the third generation quark doublet $Q$ can be constrained by the measurement of the both the dijet and the $pp\to b\bar b$ invariant mass distributions. Repeating the analysis of~\cite{Alioli:2017nzr} (which does not use $b$-tags), we find that the constraints coming from available dijet measurements~\cite{Aad:2013tea,Aad:2014vwa,Chatrchyan:2012bja} and projections at 300\,fb$^{-1}$ and 3\,ab$^{-1}$ are not competitive with the limit obtained in this paper from the $t\bar t$ invariant mass distribution. Ref.~\cite{Aaboud:2016jed} uses $b$-tagging to measure the $b\bar b$ production cross section at 7\,TeV center of mass energy. Given the limited energy and the limited invariant mass range that is explored ($m_{b\bar b}<1$\,TeV), and taking into account the large size of both systematic and theoretical uncertainties (due to the absence of full NNLO calculations for bottom production), we expect the constraints on operators involving the third generation quark doublet $Q$ coming from Ref.~\cite{Aaboud:2016jed} to be subleading to the one derived here from the $t\bar t$ distribution.

\begin{figure}[t]
\begin{center}
~~\includegraphics[width=0.475\textwidth]{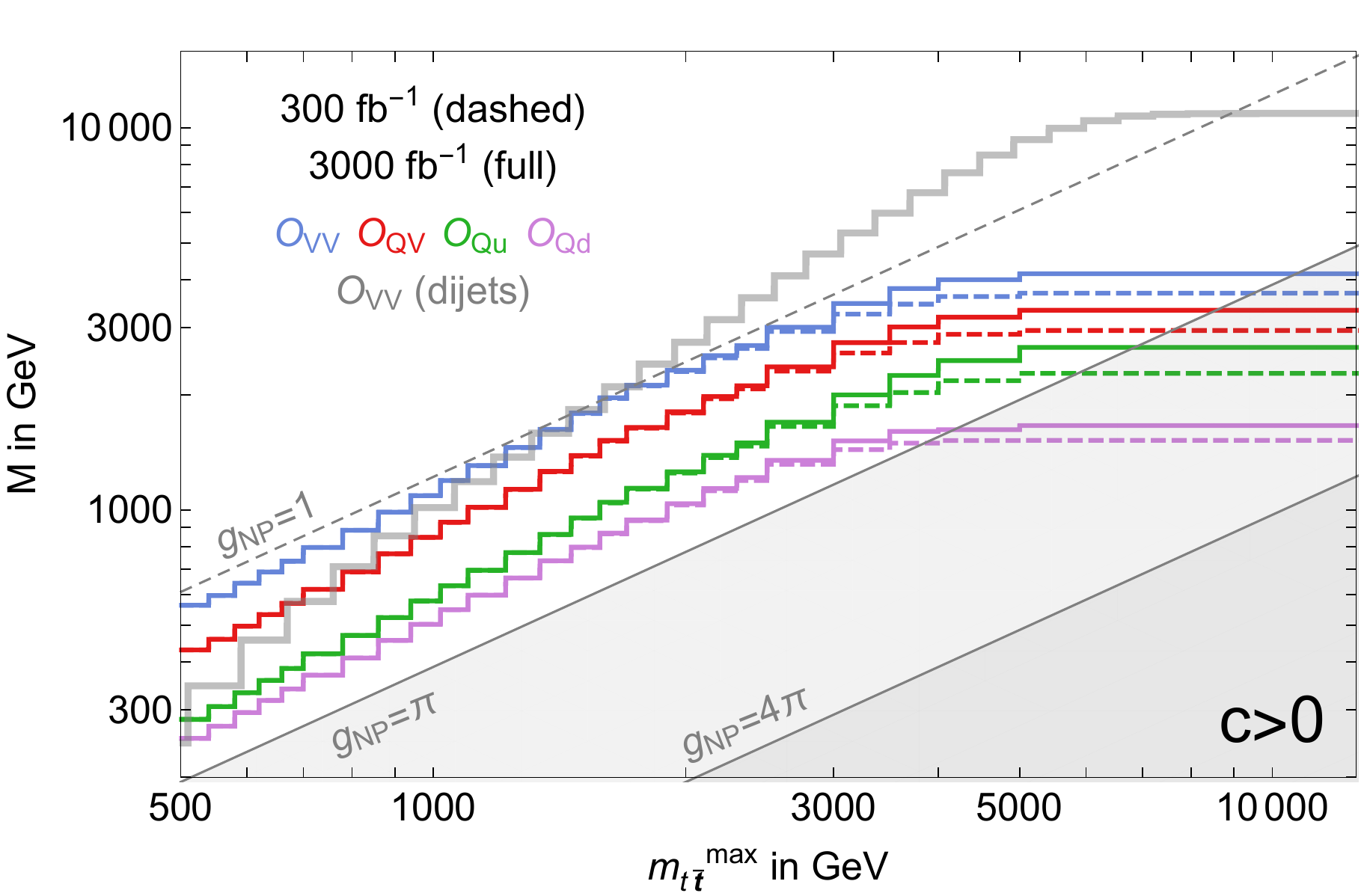}~~~~~~~\includegraphics[width=0.475\textwidth]{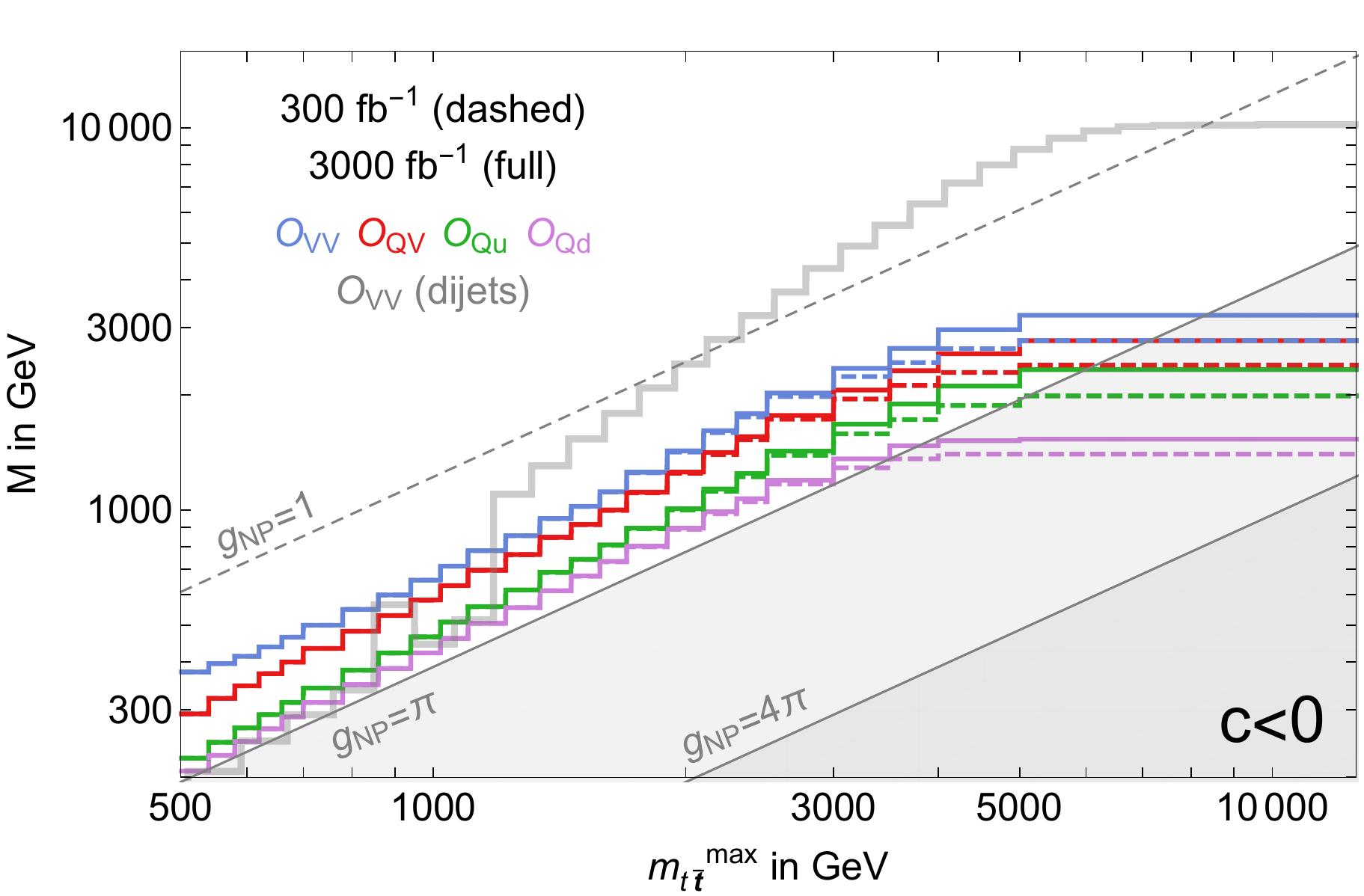}
\end{center}
\vspace{-.3cm}
\caption{ \footnotesize
Projected 95\% CL bounds on four fermion operator mass scales. We show the dependence of the bounds on the $t\bar t$ invariant mass cut $\mttcut$. On the left the bounds are shown for the case of positive operator coefficients, corresponding to constructive interference with the SM amplitude, while on right we show the case of negative operator coefficients and destructive interference. We show the estimated region of validity of the EFT description (above the straight lines) for various assumptions about the coupling strength in the short distance model generating the effective operators.
}
\label{mttcut}
\end{figure}

While we can bound the size of the operator coefficients without knowing about the physics that generates such operators, the validity of the bounds we obtain depends on such details. The reason is clear: if Eq.~(\ref{deltaL4f}) is obtained integrating out some state of mass $m_{NP}$, Eq.~(\ref{deltaL4f}) cannot properly describe the $t\bar t$ invariant mass distribution for $m_{t\bar t}$ above $m_{NP}$. 

Assuming Eq.~(\ref{deltaL4f}) is obtained by integrating out, at tree level, states of mass $m_{NP}$ coupled with strength $g_{NP}$, one would approximately expect $M\sim (g_S/g_{NP})m_{NP}$. This implies that validity of the EFT description requires
\be\label{eftvalidity}
M(\mttcut\sim m_{NP})\gtrsim\frac{g_s}{g_{NP}} m_{NP} \, \, .
\ee
We display such limits in Fig.~(\ref{mttcut}) for various values of $g_{NP}$. It should be stressed that Eq.~(\ref{eftvalidity}) is by no means rigorous and the exact regime of validity of the EFT description can only be evaluated by calculating the $t \bar t$ invariant mass distribution within a complete model and comparing to the EFT prediction.

To conclude this section we would like to point out another aspect of the bounds we described. Even though the operators in Table~(\ref{tab4f}) can interfere with the SM $q\bar q\to t\bar t$ amplitude, the bounds we obtain correspond to parameter points with similar contributions from the interference and the quadratic term in Eq.~(\ref{sigmaeft}). As an example, the projected bound on $c_{VV}$ at 300\,fb$^{-1}$ changes from $[-3.9,2.2]\times 10^{-3}$ to $[-4.0,4.0]\times 10^{-3}$ by dropping quadratic terms in the amplitude. In this situation, one possible concern is the presence of operators of dimension 8, which we have not taken into account, potentially affecting our analysis.

To address this concern, let us again consider the situation in which NP of mass $m_{NP}$ and coupling $g_{NP}$ has been integrated out to obtain Eq.~(\ref{deltaL4f}). Let us also imagine operators of dimension 8 are generated at the same time. Corrections to the $t\bar t$ cross section, for $m_{t\bar t}\approx E$, are approximately given by
\be\label{dim8}
\frac{\delta\sigma_{t\bar t}}{\sigma^{SM}_{t\bar t}}\sim \frac{g_{NP}^2}{g_s^2}\frac{E^2}{m_{NP}^2}+\frac{g_{NP}^4}{g_s^4}\frac{E^4}{m_{NP}^4}+\frac{g_{NP}^2}{g_s^2}\frac{E^4}{m_{NP}^4}+\ldots\, \, .
\ee
The terms on the right hand side of Eq.~(\ref{dim8}) represent, from left to right, SM interference with dimension 6, dimension 6 squared, and SM interference with dimension 8. Under the assumption that $E\lesssim m_{NP}$, the third term never dominates over the second if $g_{NP}\gtrsim g_s$. This mild strong coupling requirement is also the region of parameter space where our bounds are most relevant, as suggested by Fig.~(\ref{mttcut}) and Eq.~(\ref{eftvalidity}).


\section{Implications}
\label{sec:implications}

 As our analysis in the previous section shows, our bounds are relevant for models with heavy new states, $m_{NP}\gg m_t$, with moderately large couplings, $g_{NP}\gtrsim g_s$. We now discuss motivated examples where these two features are realized.
 

\subsection{Partially composite tops}

Composite Higgs models~\cite{Contino:2010rs,Panico:2015jxa} stand out as particularly relevant for our bounds as they predict new heavy states sharing sizable interactions and mixings with SM states. The mechanism through which fermion masses are generated in these models, partial compositeness, implies that one or both helicities of the top quark mix strongly with resonances from the composite sector.

At low energy this leads to a particular power counting for the four fermions operators that are generated. Assuming for simplicity that the right-handed helicity of the top quark is a composite state, up to order one factors,
\be\label{comptr}
\Delta \mathscr L \approx \frac{g_\rho^2}{m_\rho^2} P\left(\frac{g_{SM}}{g_\rho}\psi,t_R\right)\,,
\ee
where $P$ is a gauge and Lorentz invariant polynomial of degree four, $m_\rho$ is the mass of the composite states, $g_\rho$ represents their typical interaction strength, with $g_{SM}\lesssim g_\rho\lesssim 4\pi$.  Finally $g_{SM}\sim 1$ could represent one of the SM gauge couplings or the top Yukawa coupling $y_t$. Additional power counting rules can be found in Ref.~\cite{Panico:2015jxa}. 

A toy model realizing Eq.~(\ref{comptr}) is shown in Appendix~(\ref{partialcomp}). In that example the mass of a massive gluon, $m_{\mathcal G}$, can be identified with $m_\rho$, and its coupling to composite states, $g_{\mathcal G}$, can be identified with $g_\rho$.
Up to $O(1)$ factors, $O_{UV}$ is generated with coefficient $c_{UV}\sim m_t^2/m_{\mathcal G}^2$.\footnote{Analogously to Eq.~(\ref{opscombo}) we define $O_{UV}\equiv (\bar U\gamma^\mu T^AU)(\bar q\gamma_\mu T^Aq+\bar u\gamma_\mu T^Au+\bar d\gamma_\mu T^Ad)$. Since no angular information is used to extract the bounds, constraints on $O_{UV}$ and $O_{QV}$ (Figs.~(\ref{barplot}) and~(\ref{mttcut})) are equivalent.} According to Fig.~(\ref{mttcut}), the projected bounds at 300\,fb$^{-1}$ imply $m_{\mathcal G}\gtrsim 3$\,TeV\@. Given that $g_{NP}\sim g_s$,  this is marginally consistent with EFT validity.

Other dimension 6 operators also appear. There is a contribution to $\textrm{Z}\sim (g_s^2/g_{\mathcal G}^2)(m_W^2/m_{\mathcal G}^2)$, so that the same model can be constrained by the dijet analysis of Ref.~\cite{Alioli:2017jdo}. This constraint is negligible for $g_{\mathcal G} \gtrsim 3-4$ (see Fig.~(\ref{mttcut})).

The four top operator 
\be\label{fourtr}
-\frac{g_{\mathcal G}^2}{2 m_{\mathcal G}^2}\left(\bar t_R \gamma_\mu T^A t_R\right)^2
\ee
is also generated and the size of its coefficient is enhanced for large $g_{\mathcal G}$.
A bound on the coefficient of this operator corresponding to $m_{\mathcal G}/g_{\mathcal G} > 0.35$\,TeV was extracted by Ref.~\cite{Zhang:2017mls} by using the upper bound on the $pp\to t\bar t t \bar t$ cross section from CMS~\cite{Sirunyan:2017uyt}. Given the limit $m_{\mathcal G}\gtrsim 3$\,TeV from $O_{UV}$, the four top measurement is a subleading constraint when $4\lesssim g_{\mathcal G}\lesssim 10$.  In this regime, measurements of the $t\bar t$ differential cross section are the leading constraint on the model.


\subsection{Flavor models}
Given the strong constraints that exist on four fermion operators involving only the light generations of quarks~\cite{Alioli:2017jdo}, bounds coming from $t\bar t$ production will be relevant only if some degree of flavor non-universality enhances third generation couplings.
In this framework, flavor violation can remain under control if the underlying NP model respects flavor symmetries that the EFT then inherits.

One possibility is that the EFT respects Minimal Flavor Violation (MFV)~\cite{DAmbrosio:2002vsn}, such that the SM Yukawas, $Y_{U,D}$, are the only flavor violating spurions that enter effective operators. In this setup all of the operators in Table~(\ref{tab4f}) can be generated by taking the product of one flavor singlet current and one of the following two bilinears
\be\label{mfvbil}
\bar q Y_U Y_U^\dagger \gamma^\mu T^A q,\quad \bar u Y_U^\dagger Y_U \gamma^\mu T^A u  \, \, ,
\ee
where $q$ and $u$ are now three dimensional vectors in flavor space. The currents in Eq.~(\ref{mfvbil}) are singlets under the SM flavor group (because $Y_U$ transforms as a $\mathbf{3} \otimes \mathbf{\bar 3}$ under $U(3)_q \otimes U(3)_u$).  Neglecting light quark masses, $Y_U Y_U^\dagger=Y_U^\dagger Y_U=y_t^2\delta_{i3}\delta_{j3}$.
Generic UV completions will also generate flavor universal operators that are the product of two flavor singlet currents.  The bounds from dijets on flavor universal operators are a factor of $\sim10$ stronger than the bounds on operators including tops (see for example Fig.~(\ref{mttcut})).  Bound from tops can still be relevant if a mild tuning suppresses the flavor universal operators.

\begin{figure}[t]
\begin{center}
~~\includegraphics[width=0.475\textwidth]{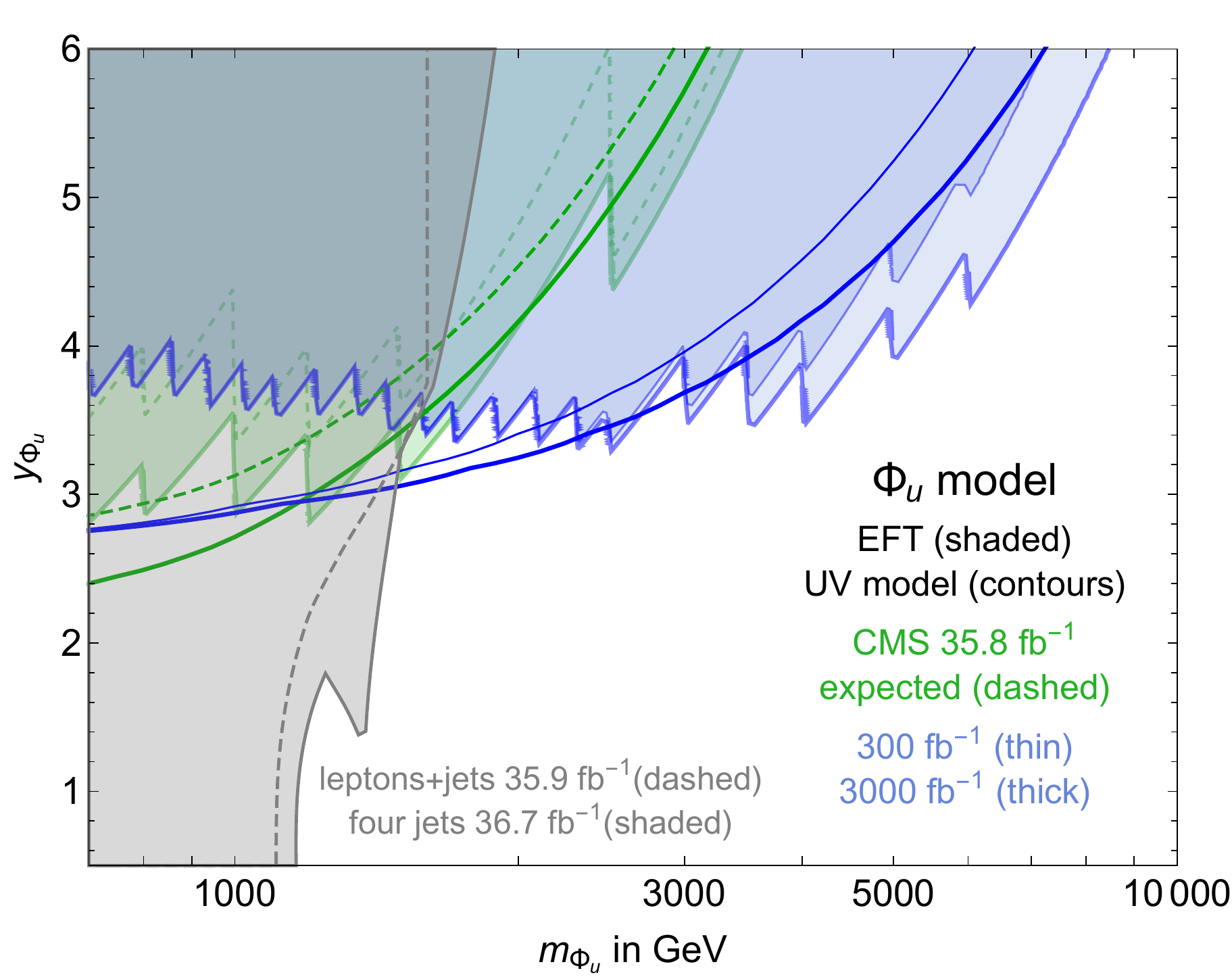}~~~~~~~\includegraphics[width=0.475\textwidth]{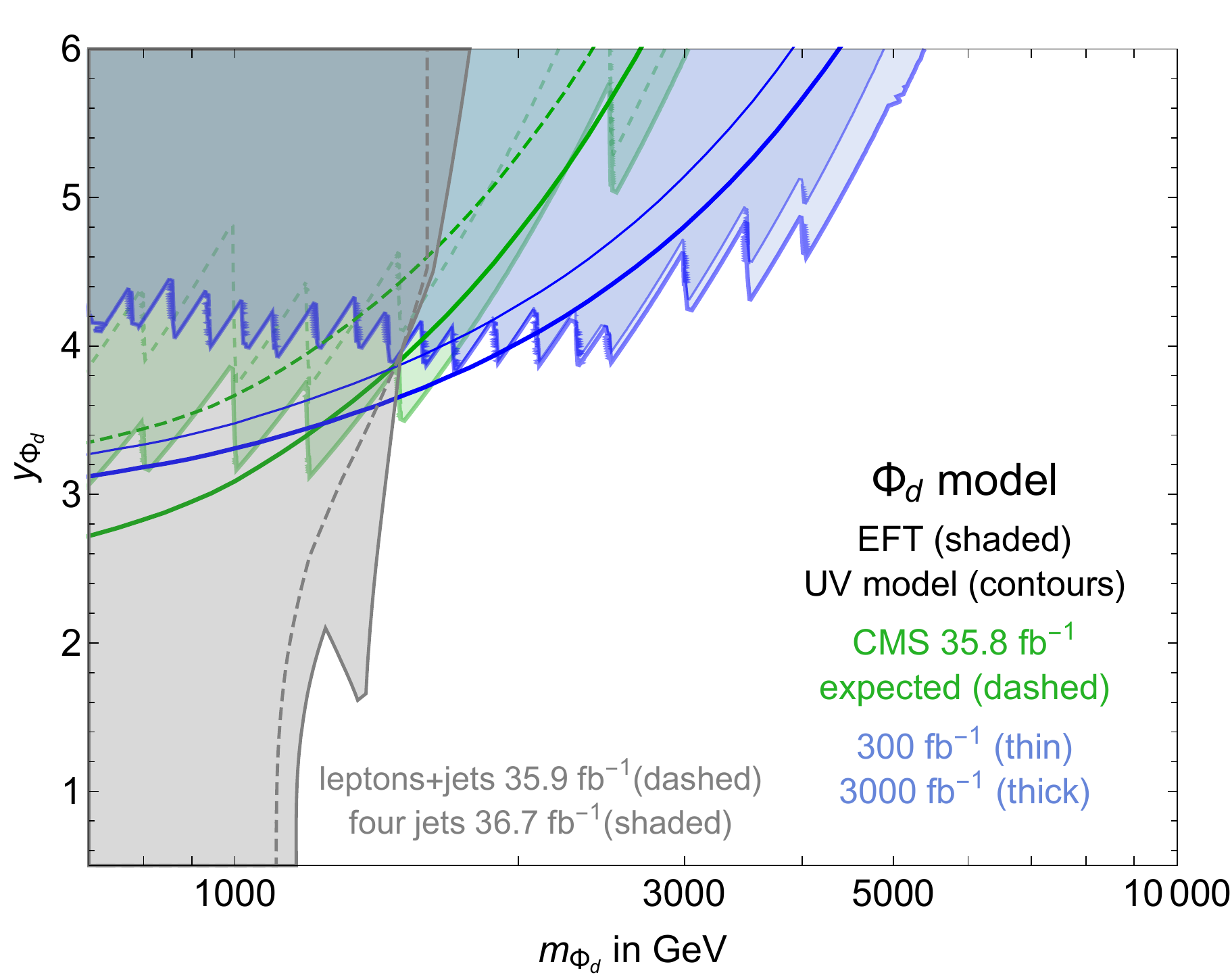}
\end{center}
\vspace{-.3cm}
\caption{ \footnotesize
\emph{Left}: constraints on the mass and coupling of a color octet electroweak doublet with hypercharge $Y=-1/2$, transforming as a doublet under the $U(2)_u$ flavor group acting on the first two generation right-handed up quarks, see Eq.~(\ref{phiu}). \emph{Right}: constraints on a color octet electroweak doublet with hypercharge $Y=1/2$, this time coupling to the first two generation right-handed down quarks with a $U(2)_d$ invariant coupling. We show 95\% CL constraints from CMS measurements of the $t\bar t$ invariant mass distribution~\cite{Sirunyan:2018wem} and high luminosity projections.
Shaded regions show the constraints extracted from the low energy EFT description as in Eq.~(\ref{eftphiu}), while solid contours show the bounds obtained by calculating the corrections to the $t\bar t$ differential cross section using the full model.  Expected CMS exclusions are also displayed (green dashed contours).  The dashed gray contours show the bounds on direct pair production of $\Phi_{u,d}$ followed by decay to top plus jet~\cite{Sirunyan:2017yta}, while the shaded gray regions show the bounds on pair production of $\Phi_{u,d}$ followed by decay to bottom plus jet~\cite{Aaboud:2017nmi}.
}
\label{scalarmodel}
\end{figure}

Alternatively, operators with tops can naturally dominate if flavor violation respects a reduced symmetry group such as $U(2)^3$~\cite{Barbieri:2011ci}. In this case it is straightforward to identify UV completions where only operators such as those in Table~(\ref{tab4f}) are generated.

As an example we extend the SM by including a complex color octet scalar $\Phi_u$ of mass $m_\Phi$.  We take $\Phi_u$  to be a doublet under $SU(2)_{EW}$  with hypercharge $Y=-1/2$, transforming as a doublet under the flavor $U(2)_u$ corresponding to the  right-handed up type quarks from the first two generations. We consider the following interactions
\be\label{phiu}
 y_{\Phi_u} \,\Phi_u^A\,\bar Q T^A u +{\textrm{h.c.}}\, \, .
\ee
It is straightforward to vary the quantum numbers of the scalar mediator, so that it can couple to different quark bilinears. Integrating out $\Phi_u$ leads to a single four fermion operator in the low energy theory,
\be\label{eftphiu}
\Delta \mathscr L = \frac{y^2_{\Phi_u}}{m_\Phi^2} \left(\bar Q T^A u\right)\left( \bar u T^A Q\right) = -\frac{1}{6}\frac{y^2_{\Phi_u}}{m_\Phi^2} \left(\bar Q \gamma^\mu T^A Q\right)\left( \bar u \gamma_\mu T^A u\right)+\frac{2}{9}\frac{y^2_{\Phi_u}}{m_\Phi^2} \left(\bar Q \gamma^\mu Q\right)\left( \bar u \gamma_\mu  u\right)  \, \, .
\ee
In the last equality we use color and Lorentz Fierz identities to bring the operator to the canonical form used in Table~(\ref{tab4f}). Both octet and singlet color structures are generated.

This model is particularly interesting in relations to the bounds we have derived, since while the scalar $\Phi_u$ cannot be resonantly produced it can contribute to the $pp\to t\bar t$ differential cross section.

Bounds on $\Phi_u$ are shown in the left panel of Fig.~(\ref{scalarmodel}). Constraints from measurements of the $t\bar t$ invariant mass distribution are extracted both in the full model and by using the EFT description of Eq.~(\ref{eftphiu}).  For the EFT bounds we fit to masses $m_{t \bar t} < m_\Phi$ to ensure validity of the EFT description (the jaggedness of the EFT bounds results from the binning of the $m_{t \bar t}$ spectrum).  We find approximate agreement between the bound using the EFT and the full model, when $m_\Phi \gtrsim 2$\,TeV, verifying that operators of dimension larger than 6 do not play an important role in this regime.  For lighter masses, the EFT gives a weaker bound than the full model because fitting to the low energy subset, $m_{t \bar t} < m_\Phi$, is conservative.

We compare the above bounds from the $t \bar t$ invariant mass spectrum to bounds from direct pair production of $\Phi_u$.  Bounds on pair production of the neutral component of $\Phi_u$ followed by its decay to $\bar tu(t\bar u)$ are extracted from Fig.~(3) of Ref.~\cite{Sirunyan:2017yta}, which uses 35.9\,fb$^{-1}$ at 13\,TeV\@. For pair production of the charged component of $\Phi_u$ followed by decay to $\bar bu(b\bar u)$ bounds are extracted from the coloron model in Fig.~(9) of Ref.~\cite{Aaboud:2017nmi}, which uses 36.7\,fb$^{-1}$ at 13\,TeV\@. In both cases we roughly adapt bounds by neglecting a possible order one difference in acceptance between the simplified model used by the experimental collaboration and the model of Eq.~(\ref{phiu}). 
While at low masses and couplings the bounds are dominated by $\Phi_u$ pair production and decay, at larger masses and moderate to large couplings the constraint from the $t\bar t$ invariant mass distribution dominates.

Bounds on an analogous model in which a scalar $\Phi_d$ couples to light right-handed down quarks through $y_{\Phi_d} \,\Phi_d^A\,\bar Q T^A d$ are shown in the right panel of Fig.~(\ref{scalarmodel}).


\section{Conclusions}
\label{sec:conclusions}

Measurements of the $t\bar t$ invariant mass distribution at high energies, together with the high precision calculation of the $t\bar t$ cross section, significantly constrains the top quark sector of the SM EFT\@.
In this paper we use the most recent data from the CMS collaboration with a luminosity of 35.8\,fb$^{-1}$,  and NNLO QCD and NLO EW calculations of the $t\bar t$ differential cross section, to constrain dimension 6 four fermion operators modifying the shape of the $t\bar t$ invariant mass distribution at high energies. Our results are summarized in Table~(\ref{tab4f}) and Figs.~(\ref{barplot}) and~(\ref{mttcut}).
In terms of the mass scale defined in Eq.~(\ref{massscale}), the current CMS bound points to $M\gtrsim 1$\,TeV for the fully chiral operators in Table~(\ref{tab4f}) and $M\gtrsim 2$\,TeV for the combinations defined in Eq.~(\ref{opscombo}). The strongest current bound, $M\gtrsim 5.5$\,TeV, is obtained for the fully vectorlike operator $O_{VV}$, with a positive coefficient and constructive interference with the SM\@.   This bound is stronger than expected due to the downward fluctuation in the CMS data that can be observed in Fig.~(\ref{CMSvstheory}). Projected bounds at high luminosity are stronger, of order 2\,TeV and 4\,TeV for the fully chiral operators and the operator combinations, respectively.

Our bounds are applicable to NP scenarios  in which the states generating the effective operators are heavy and moderately strongly coupled. We project that with more luminosity, measurements of differential tops will be sensitive to composite Higgs models with partial compositeness in which the underlying strong sector delivers resonances with moderately large couplings, $g_\rho\gtrsim 3-4$.

If we compare our results with those obtained from other hadronic observables like Drell-Yan~\cite{Farina:2016rws} and dijets~\cite{Alioli:2017jdo} we see (for instance from Fig.~(\ref{CMSvstheory})) that for $t\bar t$ observables experimental systematics are a limiting factor. It would be worthwhile to explore statistical procedures that provide alternatives to unfolding~\cite{DAgostini:1994fjx} (for example see Ref.~\cite{cranmer_kyle_2017_1013926}), where systematic uncertainties may take a different size.

It has been shown that measurements of top quark pair differential distributions can be used to constrain the gluon PDF~\cite{Czakon:2016olj}.  There is a risk that nonzero operators from the SM EFT may bias future PDF fits, and that the resulting PDFs may lead to incorrect bounds on the size of these operators.  It would be interesting to explore the interplay of PDF fits and the SM EFT, such as the possibility of using differential top measurements to perform simultaneous fits to EFT operators and PDFs.

Finally, as a next step it would be interesting to perform a multidimensional kinematic fit that goes beyond this study by including angular observables.


\section*{Acknowledgments}
We thank Kyle Cranmer, Otto Hindrichs, Alex Mitov, and Juan Rojo for helpful conversations.  CM and JTR acknowledge the CERN theory group for hospitality while part of this work was completed. MF is supported by the NSF grant PHY-1620628. CM is supported by the James Arthur Graduate Fellowship. JTR is supported by the NSF CAREER grant PHY-1554858. 
\newpage
\appendix



\section{Operators and amplitudes}\label{operators}
While the operators in Table~(\ref{tab4f}) are the only ones contributing to $pp\to t\bar t$ at leading order and neglecting SM particle masses, many more dimension 6 operators contribute to this process beyond leading order. Restricting our attention to four fermion operators we have in addition the following structures:
\begin{align}\nonumber
\widetilde{\mathcal{O}}_{Qq}^{(3)} & = \bar{Q}\gamma_\mu \tau^i Q\,\bar{q}\gamma^\mu \tau^i q\, \, , \\\nonumber
\widetilde{\mathcal{O}}_{Qq} & = \bar{Q}\gamma_\mu Q\,\bar{q}\gamma^\mu q \,\, ,\\\nonumber
\widetilde{\mathcal{O}}_{Ud} & = \bar{U}\gamma_\mu U\,\bar{d}\gamma^\mu d \,\, , \\
\widetilde{\mathcal{O}}_{Uu} & = \bar{U}\gamma_\mu U\,\bar{u}\gamma^\mu u \, \, ,\label{noint} \\\nonumber
\widetilde{\mathcal{O}}_{Uq} & = \bar{U}\gamma_\mu U\,\bar{q}\gamma^\mu q \, \, ,\\\nonumber
\widetilde{\mathcal{O}}_{Qd} & = \bar{Q}\gamma_\mu Q\,\bar{d}\gamma^\mu d\,\, , \\\nonumber
\widetilde{\mathcal{O}}_{Qu} & = \bar{Q}\gamma_\mu Q\,\bar{u}\gamma^\mu u\, \, .
\end{align}
The difference between these operators and those in Table~(\ref{tab4f}) is the color structure: all the operators in Eq.~(\ref{noint}) display a color singlet contraction and for this reason they do not interfere with the QCD amplitude of the SM\@.

The set of operators contributing to the $t\bar t$ invariant mass distribution is not exhausted by four fermion operators. The full list can be found in Ref.~\cite{Buckley:2015lku}. Again interference with the SM QCD amplitude at high energies is either vanishing or suppressed~\cite{contino}.

For the operators in Table~(\ref{tab4f}) and with the normalization of Eq.~(\ref{deltaL4f}), we can write the squared matrix element for $q(p_1)\bar q(p_2)\to t(p_3)\bar t(p_4)$ (summed over final state quantum numbers and averaged over initial state ones) as
\be
\overline{|\mathcal M|^2} =\frac{4}{9}g_s^4\left(\mathcal A_{SM}+\sum_i c_i \mathcal I_{i}+\sum_{i,j} c_ic_j \mathcal Q_{i,j}\right).
\ee
The SM $u\bar u\to t\bar t$ and $d\bar d\to t\bar t$ amplitudes are mediated by the exchange of a gluon in the $s$-channel,
\be
\mathcal A^{u,d}_{SM} =\frac{t^2+u^2+4 m_t^2 s-2m_t^4}{s^2} \, \, ,
\ee
where as usual $s=(p_1+p_2)^2$, $t=(p_1-p_3)^2$, and $u=(p_1-p_4)^2$.
The interference terms read
\begin{align}
\mathcal I^{u}_{Qq^{(3)}}=-\mathcal I^{d}_{Qq^{(3)}}=\mathcal I^{u,d}_{{Qq}}=\mathcal I_{{Uu}}=\mathcal I_{{Ud}} &= \frac{u^2-m_t^2(3 u+ t)+3 m_t^4}{s m_t^2}\, \, ,\\
\mathcal I^{u,d}_{Uq}=\mathcal I_{{Qu}} &=\frac{t^2-m_t^2(3 t+ u)+3 m_t^4}{s m_t^2}\, \, .
\end{align}
Finally the quadratic terms are reported  in the following two tables
\begin{center}
{\small
\begin{tabular}{c||ccccc} 
 $\mathcal Q^{(u)}$ & $\tilde O_{Qq}$ &$O_{Qq}$ &$O_{Qu}$ &$O_{Uq}$ & $O_{Uu}$  \\ \hline \hline 
$\tilde O_{Qq}$ &$\tfrac{1}{2}Q(u)$&$Q(u)$&0&$m_t^2 s$&0\\
$O_{Qq}$ &&$\tfrac{1}{2}Q(u)$&0&$m_t^2 s$&0\\
$O_{Qu}$ &&&$\tfrac{1}{2}Q(t)$&0&$m_t^2 s$\\
$O_{Uq}$ &&&&$\tfrac{1}{2}Q(t)$&0\\
$O_{Uu}$ &&&&&$\tfrac{1}{2}Q(u)$\\
\end{tabular}\\ \vspace{0.5cm}
\begin{tabular}{c||ccccc} 
 $\mathcal Q^{(d)}$ & $\tilde O_{Qq}$ &$O_{Qq}$ &$O_{Qd}$ &$O_{Uq}$ & $O_{Ud}$  \\ \hline \hline 
$\tilde O_{Qq}$ &$\tfrac{1}{2}Q(u)$&$-Q(u)$&0&$-m_t^2 s$&0\\
$O_{Qq}$ &&$\tfrac{1}{2}Q(u)$&0&$m_t^2 s$&0\\
$O_{Qd}$ &&&$\tfrac{1}{2}Q(t)$&0&$m_t^2 s$\\
$O_{Uq}$ &&&&$\tfrac{1}{2}Q(t)$&0\\
$O_{Ud}$ &&&&&$\tfrac{1}{2}Q(u)$\\
\end{tabular}

} 
\end{center}
where $Q(x)=(x/m_t^2-1)^2$. The tables are symmetric and lower diagonal elements are not shown.


\section{Toy model for a composite $t_R$}\label{partialcomp}

In this section we introduce a toy model with a composite right-handed top quark, realizing the power counting in Eq.~(\ref{comptr}). This model does not solve the hierarchy problem but can be easily extended to do so~\cite{Panico:2011pw,Matsedonskyi:2012ym,Panico:2012uw,DeSimone:2012fs}.

On top of the particles and interactions already present in the SM, we introduce a massive $SU(3)$ color octet vector $\mathcal G$ (an excited state of the SM gluon), and an $SU(2)$ singlet vectorlike quark $T$ with hypercharge $2/3$.

We consider the following Lagrangian for $\mathcal G$ and $T$:
\begin{align}
\Delta \mathscr L =& -\frac{1}{4}D_{[\mu}\mathcal G_{\nu ]}^{A} D^{[\mu}\mathcal G^{\nu ]A}+\frac{ m^2_{\mathcal G}}{2}\mathcal G_\mu^A\mathcal G^{A\mu}+ \bar T(i\slashed{D}-m_T) T \\
&+c_{g}\frac{g_s}{2 g_{\mathcal G}} D_{[\mu}\mathcal G_{\nu ]}^{A}  G^{\mu\nu A}-g_{\mathcal G} \mathcal G^A_{\mu} \bar T\gamma^\mu T^A T\\
&+y_1 \bar Q_L \tilde H t_R+y_2 \bar Q_L \tilde H T_R+m\bar T_L t_R+{\textrm{h.c.}}\, \, .
\end{align}
Integrating out $\mathcal G$ and $T$, and using  equations of motion of the light fields, yields the following low energy Lagrangian 
\begin{align}
\Delta \mathscr L =&\left(y_1+y_2\frac{m}{m_T}\right)\bar Q_L \tilde H t_R \\
&-\frac{1}{2m_{\mathcal G}^2}\left(c_g\frac{g_s^2}{g_{\mathcal G}} J_\mu^A+ g_{\mathcal G}\frac{m^2}{m_T^2}\bar t_R \gamma_\mu T^A t_R\right)^2 \, \, ,
\end{align}
where $J_\mu^A$ is the fermionic current of gauged $SU(3)$. We assumed $g_s\ll g_{\mathcal G}$ and $m\lesssim m_T$. If $c_g\sim 1$ and $m\sim m_T$, we obtain the same power counting of Eq.~(\ref{comptr}) after identifying $m_\rho \equiv m_{\mathcal G}$ and $g_\rho \equiv g_{\mathcal G}$.

\pagestyle{plain}
\bibliographystyle{jhep}
\small
\bibliography{biblio}

\end{document}